\documentclass[superscriptaddress,twocolumn,nofootinbib]{revtex4}
\usepackage{graphicx}
\usepackage{color}
\usepackage{enumerate}
\usepackage{amssymb}
\usepackage{amsmath}
\usepackage{amsfonts}
\usepackage{bbm}
\usepackage{dsfont}
\def\be{\begin{equation}} 
\def\ee{\end{equation}}

\def\uf{$^{236}$U}
\def\ut{$^{235}$U}
\def\Hb{H_{\rm bridge}}
\def\unedf1{\textsc{unedf1}}
\def\Eex{E_{\rm ex}}
\def\inn{{\rm in}}
\def\Ts{\mathcal{T}}

\begin{document}

\title{Modeling barrier-top fission dynamics in a
discrete-basis formalism}

\author{G.F. Bertsch}
\affiliation{ 
Department of Physics and Institute for Nuclear Theory, Box 351560, 
University of Washington, Seattle, Washington 98915, USA}

\author{K. Hagino}
\affiliation{ 
Department of Physics, Kyoto University, Kyoto 606-8502,  Japan} 

\begin{abstract}
A configuration-interaction model is presented for the barrier
region of induced fission.  
The configuration space is composed 
of seniority-zero configurations constructed from self-consistent
mean-field wave functions. The Hamiltonian matrix elements between
configurations include diabatic and pairing interactions between
particles. Other aspects of the Hamiltonian 
are treated statistically,
guided by phenomenological input of compound-nucleus transmission 
coefficients. In this exploratory study the configuration space
is restricted to neutron excitations only. A key observable 
calculated in the model is the fission-to-capture branching ratio.
We find that both pairing and diabatic interactions are important
for achieving large branching to the fission channels.  
In accordance with the transition-state theory of fission, 
the calculated branching ratio is found to be quite insensitive to the
fission decay widths of the pre-scission configurations.
However, the barrier-top dynamics appear to be 
quite different from transition-state theory in that the 
transport is distributed over many excited configurations at the
barrier top.\\
\end{abstract}

\maketitle

\section{Introduction}
The theory of fission at a barrier top energies has been one of 
the few topics in low-energy nuclear physics that has been beyond the
purview of the configuration-interaction (CI) framework of modern
nuclear theory.  In that framework one builds a matrix 
Hamiltonian in a space of Slater determinants composed of nucleon orbitals, with
matrix elements derived from nucleon-nucleon interactions.
In this work we construct a CI model of fission dynamics
with parameters  guided by our present knowledge of the nuclear Hamiltonian.  
From a computational point of view, this formulation has some of the ingredients of the
Generator Coordinate Method (GCM) which has also been applied
to fission theory \cite{re16}.  However, the GCM method treats the dynamics
as a Schr\"odinger equation of a few collective coordinates rather than
as a discrete-basis matrix Hamiltonian equation.

The present CI model 
is too simplified to provide a quantitative theory, but hopefully it is
sufficiently realistic to allow qualitative conclusions about
the fission dynamics at the barrier.  See Refs. \cite{be22,388,385,ha20} for our 
previous simplified models to that end.  While the model is realistic
in that the configurations are built from well-documented energy-density
functionals\footnote{We ignore the conceptual differences between an
energy functional and a Hamiltonian.}, that space is severely truncated, 
allowing only neutron excitations in seniority-zero configurations.
There are two types of 
residual interaction that are 
active in a seniority-zero basis, namely the pairing interaction and
an interaction associated with diabatic evolution of the wave function.  

In order to make a complete theory of reaction cross sections, the 
Hamiltonian bridge across the barrier must also be augmented 
with statistical reservoirs.  That includes 
the configurations that make up the compound nucleus and those
that link the bridge states to the final fission channels. They
will be treated in a statistical way based on the Gaussian
Orthogonal Ensemble (GOE).

The basic physical quantities to be computed are the $\mathcal{S}$-matrix 
reaction probabilities $\mathcal{T}_{{\rm in},k}$ to 
capture  or fission  
channels\footnote{These are to be distinguished from
the transmission factors $T$ between channels and the compound nucleus.  
We will use both in the present work.},
\be
\mathcal{T}_{{\rm in},k} =\sum_{j \in k}| \mathcal{S}_{{\rm in},j} | ^2.
\label{Tij}
\ee
Here ``in"  is the neutron entrance channel, and $k$ = ``cap'' or
``f''  is the set of exit channels of a given type.
The present model is not detailed
enough to calculate the absolute reaction probabilities, but
we believe it has enough microscopic input to treat the
energy dependence of $\mathcal{T}_{{\rm in},k}$ and some aspects of 
the branching ratio, defined
experimentally as
\be
\alpha^{-1} = \frac{\int dE\, \mathcal{T}_{\rm in,f}(E)}{\int dE\, 
\mathcal{T}_{\rm in,cap}(E)}
\label{alpha}
\ee
where the integral is taken over some experimentally defined energy interval.

In the next three sections below, we present the reaction theory formalism,
the construction of the bridge Hamiltonian $\Hb$, and
the results of calculations with a full Hamiltonian that links
an entrance channel to a set of exit channels.
In this paper, we only discuss the barrier-top fission of 
$^{236}$U, but the formalism is general and can be applied to other nuclei 
as well. 

\section{Reaction theory formalism}
\label{reaction-theory}
There are several ways to formulate reaction theory in a CI framework. 
The ones that we have employed
are the $\mathcal{S}$-matrix theory leading to the Datta formula \cite{da01},  
the $K$-matrix formula\cite{bertsch2000} \footnote{The $K$-matrix formalism is close
to the $R$-matrix formalism; the latter is commonly used to fit 
resonance data.}, 
and the direct solution for the  
wave function.  The methods are algebraically equivalent
\cite{al20,al21}.  
The theory requires two matrices, one for the Hamiltonian
of the internal states and 
one for its couplings to the various continuum
channels. The Hamiltonian $H$ is a real matrix of dimension
$N_\mu $ where $N_\mu$ is the number of 
configurations in the fused system. 
The other matrix is $W$, a real matrix of dimension $N_\mu \times N_{\rm ch}$ composed
of reduced-width amplitudes $W_{\mu,i}$ coupling configuration
$\mu$ to channel $i$.  
Here, $N_{\rm ch}$ is the number of channels.
The partial width to decay from the
state $\mu$ through the channel $i$ is
\be
\Gamma_{\mu,i} = 2  W_{\mu,i}^2.
\label{GammafromW}
\ee
In case the channel couples to more than one state, one needs to 
consider the full decay matrix associated with the channel,
\be
\Gamma_{\mu,\mu',i}  =  2 W_{\mu,i} W_{\mu',i}.
\ee
The basis states 
constructed by the GCM 
are not necessarily orthogonal and one also needs  the
matrix of overlaps $S$ between configurations.

In this work we do not need the $\mathcal{S}$-matrix itself, but only 
reaction 
probabilities $\mathcal{T}_{i,j}$ between one channel $i$ and another $j$, as
given in Eq. (1) above.
They can be conveniently calculated by the trace formula\footnote{An
equivalent formula has also been used in nuclear reaction theory 
\cite{yo86,co94}.},
\be
\mathcal{T}_{i,j}(E) = {\rm Tr}\,\left(  \Gamma_i  G(E)  \Gamma_j  G^\dagger(E)\right),
\label{datta}
\ee
where $G(E)$ is the Greens' function\footnote{Here we have neglected
level shifts due to the channel couplings.}
\be
 G(E) = \left(  H - i \sum_k  \Gamma_k/2 -  S E\right)^{-1}. 
\ee

\begin{figure}
\includegraphics[width=1.0\columnwidth]{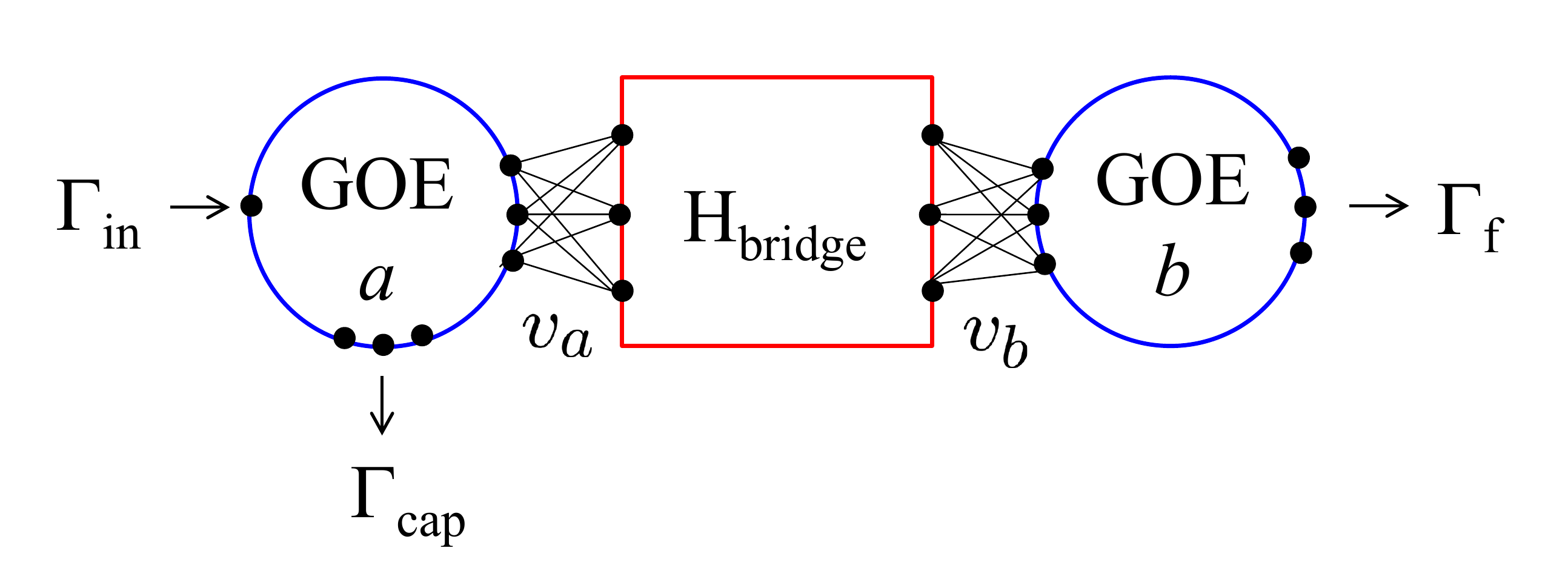}
\caption{
\label{connectivity}
Connectivity of the Hamiltonian for calculating reaction transmission
factors.   The large circles represent states of the compound nucleus
(a) and of the post-barrier configurations (b), both modeled by Hamiltonians
of the Gaussian Orthogonal Ensemble (GOE).
The rectangular box contains the bridge configurations modeled by
an explicit microscopic Hamiltonian.  The black dots
represent states that 
connect the different domains of the full
Hamiltonian. There is a single entrance channel but multiple exit
channels to bound states of the compound nucleus and to 
states that decay by fission.  Couplings to the
entrance, capture, and fission channels have associated decay widths
given by $\Gamma_{\rm in},\Gamma_{\rm cap}$ and $\Gamma_f$, respectively.
}
\end{figure}

\section{CI model space and Hamiltonian matrix elements}   

The space of internal states is composed of three sets: those of the 
compound nucleus, those of the bridge configurations, and those 
beyond the bridge that ultimately lead to fission.
Their Hamiltonian connections are schematically
shown in Fig. 1.  The dots identify individual configurations at the
borders of different sets of states.  The circle $a$ denotes the
compound-nucleus Hamiltonian as defined by a GOE.  We also treat the configurations
beyond the barrier (circle b) statistically in the same way. Specific details 
on their definition and properties  are given in Appendix A.
The rectangular block represents the bridge states that cross the
barrier.  They are composed of configurations constructed by a 
constrained minimization procedure, as is done in the first steps of the
GCM.  The configurations are linked by parameterized nucleon-nucleon
interaction matrix elements. Details are described in 
the next section below. 

The reaction theory also requires decay-width matrices for the entrance
channel, the capture channels, and the fission channels.  They are
depicted in Fig. 1 as $\Gamma_{\rm in},\Gamma_{\rm cap}$, and $\Gamma_{\rm f}$. For 
the present model, we have good information about the first two 
widths but no quantitative information about the fission widths on
the end\footnote{See Ref. \cite{375} for a computational framework
to estimate these decay widths.}.

\subsection{Bridge Hamiltonian}
\label{IIIb}

  We wish to construct the bridge Hamiltonian $\Hb$ as realistically as
possible, recognizing that the large dimensions and the
number of configuration-interaction matrix elements require
severe compromises.  The general scheme is easy to describe.
The first step is to define a set of 
reference states along an assumed fission path.  These are
Slater determinants of nucleon orbitals calculated by constrained
density-functional theory.
Next one builds a configuration space of particle-hole excitations
on each reference state.  We call that space a $Q$-block.
Finally one computes matrix elements. 
It should be emphasized that the
Slater-determinant basis, also called a Hartree-Fock (HF) basis,  is
fundamental to the CI approach.  It has a 
certain advantage with respect to quasi-particle bases (called HFB) which
require projections to treat specific nuclei.

The bridge Hamiltonian $\Hb$ can be written as 
\be
\Hb = \sum_q H_q + \sum_{qq'} V_{qq'}. 
\ee
Here $H_q$ is the full Hamiltonian within  a Q-block and $V_{qq'}$ is the
interaction Hamiltonian between configurations in different
Q-blocks. The next section discusses the selection of reference
configurations $q$. The construction of the $H_q$ configuration
space with its diagonal and off-diagonal matrix elements is
given in the sections following that.  

The needed  computational tools for the diagonal elements of $H_q$ are available for several 
EDF's,
notably the code \textsc{Skyax} for Skyrme functionals \cite{skyax} and
the code \textsc{HFBaxial} for Gogny functionals \cite{robledo}. 
In building the reference states, the single-particle
potential is assumed to be axially symmetric with good parity.  This allows the 
orbitals as well as
the configurations to be classified by quantum numbers for 
angular momentum about the symmetry axis and parity, $K^\pi$ \cite{byr18}. 
To   determine the diagonal energies in the Hamiltonian  we separate
the tasks of setting the absolute energies $E_q$ of the reference states
and setting the excitation energies $E_{\rm ex}(q\,\mu)$ for configurations
$\mu$ within a $Q$-block,
\be
\langle q\, \mu | H_q | q \,\mu\rangle = E_q + E_{\rm ex}(q\,\mu).
\label{Hqmu}
\ee
For the present model of $H_q$, we use the Skyrme energy functional
\unedf1 \cite{unedf1} in the Skyax code. 
Notice that the effective mass for this interaction is close to unity. 
The choice is motivated by need to
reproduce physical level densities as accurately as possible.

\subsubsection{Fission path and reference configurations}
\label{PES}

\begin{figure}
\includegraphics[width=1.0\columnwidth]{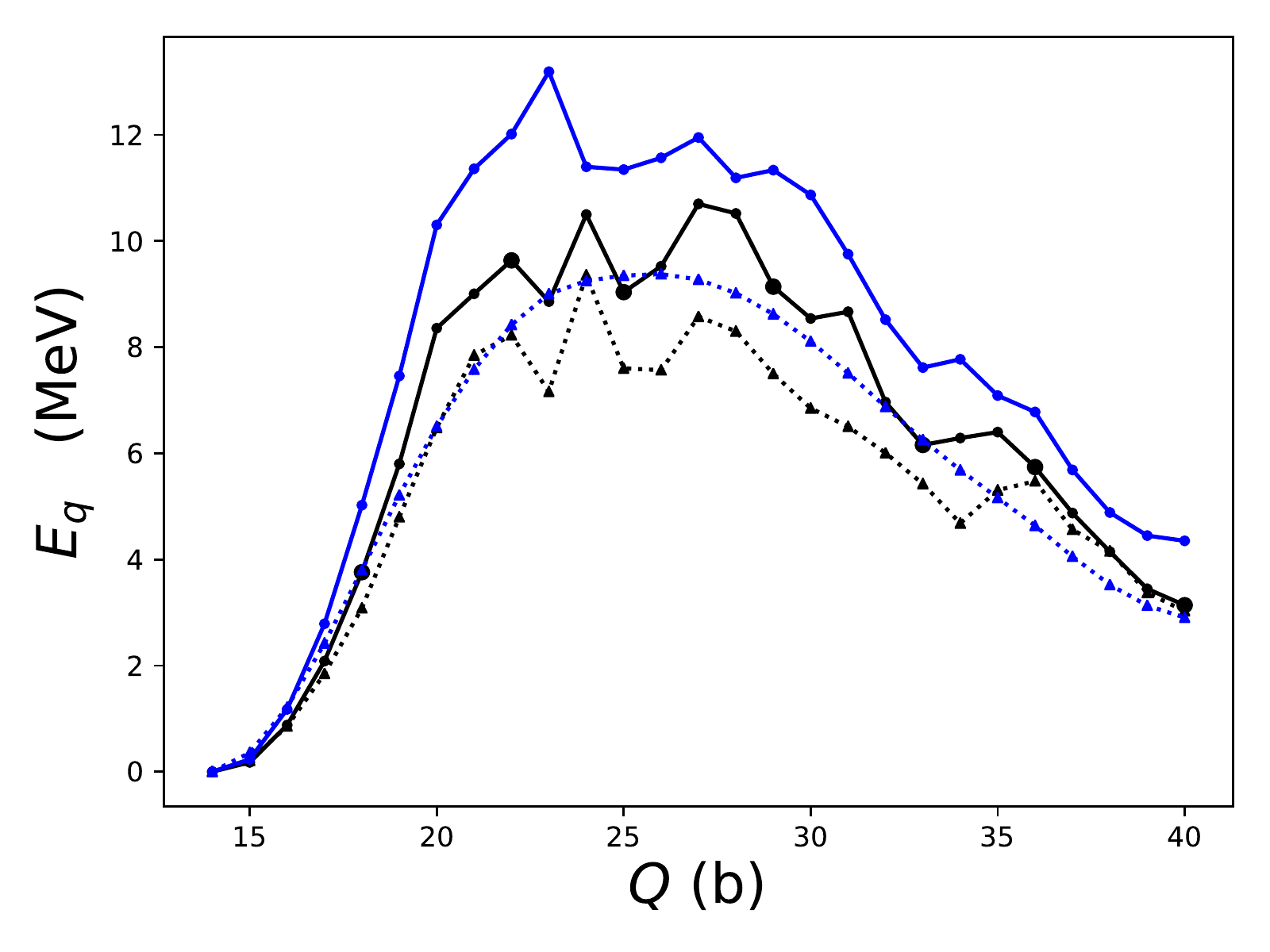}
\caption{The potential energy surface (PES) for the fission path in \uf~ as calculated 
in the HF and HFB frameworks, shown as solid and dashed lines, 
respectively. The energy functionals are the Skyrme
\unedf1 functional (black circles and triangles) and the
Gogny D1S functional (blue circles and triangles).  
\label{pes3}
}
\end{figure}

   The reference states are placed along a fission path \{$q$\} defined
by some set of constraints, as in the usual GCM.
The obvious choice is a single constraint
on the elongation of the nucleus; we use the mass quadrupole
operator\footnote{In
principle this definition can fail if the path crosses transverse
ridges \cite{du12}.} 
\be
Q = r^2 P_2(\cos\theta) \equiv z^2 - (x^2+y^2)/2.
\ee  
The reference states and associated $Q$-blocks will be labeled by an
integer $q$ set by the expectation value $\langle Q \rangle$ in units of
barns. The energy as a function of the constraint is the so-called potential
energy surface (PES).   Fig. \ref{pes3} shows a few PES plots for the 
nucleus \uf.  In our CI approach we only have discrete points
$E_q$ on the PES. In the graph the deformation ranges 
from  $q\approx14$  at the ground state minimum to $q\approx 40 $ 
near the second minimum, with the points spaced by roughly $\Delta \langle Q\rangle \approx
1$ b.  The black and blue points were calculated with the Skyrme \unedf1 
and Gogny D1S EDF's respectively.  The minimizations were carried
out in HF and HFB spaces for the circles and squares, respectively. 
Note that HF PES is far from smooth.  There
are numerous orbital crossings along the fission path and they are
responsible for abrupt changes in slope for both the Gogny and Skyrme
EDF, although the locations of the crossings differ.  
Both HF barriers are much higher than the accepted value between
5 and 6 MeV. As is well known, the
calculated barrier height is significantly lowered when the pairing
interaction is taken into account\footnote{
Triaxial deformations may also lower the barrier but they are
beyond the scope of the present model.}.  Black squares show
the Skyrme PES with neutron pairing included as described in Section \ref{vp}
below.  The lowering is not sufficient to bring the 
barrier close to the empirical value, and the PES remains
bumpy.    Ones sees a stronger decrease in barrier
height for the Gogny EDF in the HFB treatment, but it is still insufficient to be realistic.
Note that the HFB PES is quite smooth.  This
is likely an unphysical consequence of the HFB space, which
inevitably averages  over nuclei near the target one.

Since the barrier is unacceptably high we shall rescale
the reference state energies $E_q^{\rm EDF}$ 
to bring the PES closer to the empirical.  The rescaled
energies are given by 
\be
E_q  = f_{\rm pes}  E_q^{\rm EDF}.
\ee   
Here $E_q^{\rm EDF}$ is the reference energy calculated as the
difference of energies of the reference state and the
ground state at $\langle Q\rangle \approx  14 $ b.  The scaling parameter $f_{\rm pes}$
is set to $f_{\rm pes} = 0.37$ in the $\Hb$~baseline model.

The basis of states  in a GCM model need not be orthogonal.
This does not impose any conceptual difficulties for the theory
but it does add complications.
If the reference states are too close together,
the wave functions will have large overlaps and the CI 
calculational framework  becomes unstable.  On the other hand,
the reference states need to be close enough to adequately
represent the wave function at all points along the path.  
A useful measure \cite{bo90,be19a}
for setting the spacing of the reference states 
$| q\, {\rm ref} \rangle$   
is 
the quantity
$\zeta$ defined for a chain of $N$ states as
\be
\zeta  = \sum_{n = 0}^{N-1} \Delta \zeta_{n,n+1}
\label{zeta-eq}
\ee
\be
\Delta \zeta_{qq'} = (-\ln |\langle q \,{\rm ref} | q'\, {\rm ref} \rangle|)^{1/2}.  
\label{zeta-eq2}
\ee 
This assumes that 
the $K^\pi$ occupancy of the orbitals is the same all along
the chain.
It has been shown in a simplified model \cite{be22}
that spacing the states along the
chain by $\Delta \zeta \approx 1$ gives a 
fairly good approximation to the reaction probabilities.
It requires only  
5 to 6  reference states along the \uf~fission path from q=18 to q=36, and it
is large enough to neglect interactions between $Q$-blocks that are 
not nearest neighbors.

The definition Eq. (\ref{zeta-eq2}) fails 
when the occupation numbers of $K^\pi$-partitioned orbitals  
are different 
in the two configurations, in which case $\Delta \zeta=0$.  This is true
for many of the links between reference states.  For example,
we found
that  five orbital pair jumps are needed to
connect the reference configurations at each end. 
One  can still  keep $\zeta$ as a rough measure of distance by extending
the configuration space to include the particle-hole excitations
in the $Q$-blocks.
If the spaces are large enough,
all reference state  will have a partner in the neighboring $Q$-blocks. 
To determine the linking, we examine overlaps of the 
occupied orbitals in the reference configuration
with all orbitals of the same $K^\pi$ in the other $Q$-block.
The desired configuration in the second $Q$-block is the Slater determinant
of orbitals with the highest overlaps.  We call that configuration the 
diabatic partner of the reference state.  Of course the derived $\Delta \zeta$
for other configurations would vary,
but for rough studies the difference should not be important.
\begin{figure}
\includegraphics[width=1.0\columnwidth]{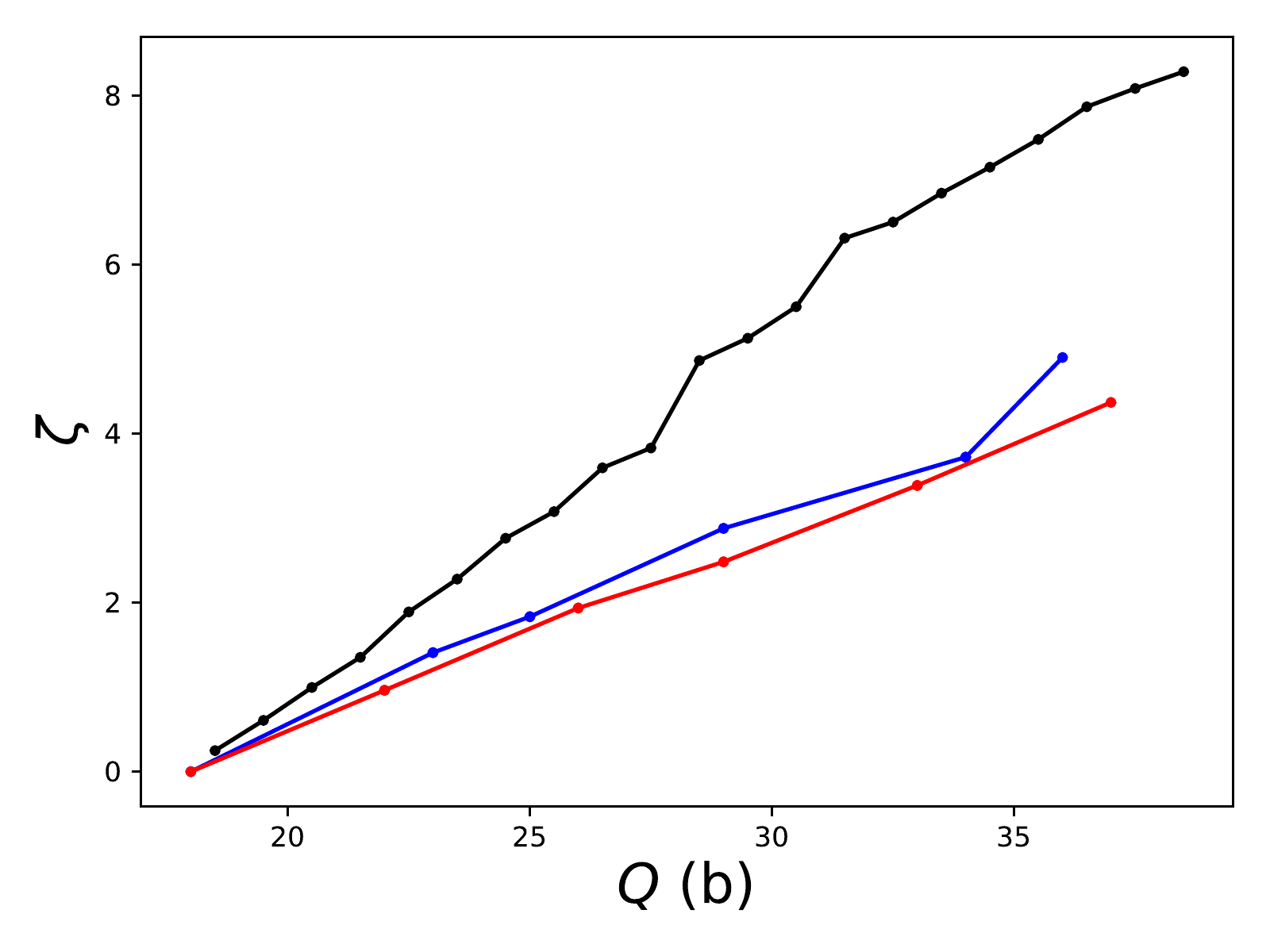}
\caption{
\label{zeta}
Overlap distances of $Q$-blocks along the fission path as defined  by Eq.
(\ref{zeta-eq}).
Black dots: a chain of 21 $Q$-blocks separated by
roughly  $\Delta Q \approx 1$ b.  Red dots: the chain
of the 6 $Q$-blocks used to construct $H_{\rm bridge}$.  Blue dots:
an alternative set of 6 $Q$-blocks  
covering about the same range of deformation.
}
\end{figure}
Fig. \ref{zeta} shows the $\zeta$ distance function across the
barrier for the \unedf1 functional with several choices for the
reference states, taking the ground state
of the left-hand configuration and the diabatic link for the
right-hand configuration. Note that the distance between the
endpoint configurations is somewhat smaller with the coarser
mesh.  This is to be expected since the finer mesh 
path gives more sensitivity to fluctuations in other degrees of
freedom.  In fact, the adiabatic prescription for defining the
path is not optimal for subbarrier fission \cite{gu14,ha20}.

For the present model, we build the $Q$-blocks on a set of 6 reference states
at deformations $q = (18,22,26, 29,33,37)$.  The configurations beyond those
on either side are assumed to be in the statistical reservoirs. 
We call this the Q6 model. In it, we assume that the 
diabatic links between the neighboring Q-blocks have the overlap 
$e^{-(\Delta\zeta)^2}$ with the overlap 
distance 
of $\Delta\zeta=1$. 
The overlaps  
between other configurations, except for those between the same configurations, 
are simply set to be zero. 

\subsubsection{$Q$-block spectrum} 
\label{Hdiag} 
The spectrum of excited configurations in a $Q$-block is generated in
the independent-particle approximation using the orbital energies
$\varepsilon_{qi}$ 
extracted from the same computer code that produced  the reference states.
The excitation energy is calculated as
\be 
E_{\rm ex}(q\,\mu) =  \sum_{i}
 \varepsilon_{qi} (\langle q\,\mu|a^\dagger_i a_i | q\,\mu\rangle -
\langle q\,{\rm ref}|a^\dagger_i a_i | q\,{\rm ref}\rangle) 
\label{E_ex}
\ee 
in an obvious notation.

Normally the occupied orbitals in the reference state are the lowest
ones in the orbital  energy spectrum, in which case $E_{\rm ex}$ is
always positive.  
In a few cases the HF minimization fails because the occupation numbers
change from one iteration to the next.  This is avoided by 
freezing the $K^\pi$ partition 
after 1500 iterations. In such cases
the converged reference state may have
one or more empty orbitals below the energy of the highest occupied orbital.  
Then
Eq.  (\ref{E_ex}) gives an unphysical negative energy.  This might be
corrected by introducing the particle-hole
interaction in the Hamiltonian.  Rather than 
complicating 
the theory this way, we
 simply ignore the sign in Eq. (\ref{E_ex}), keeping  few
with negative energy.  This is equivalent to redefining the reference
configuration in the PES as the one with the lowest energy in Eq.
(\ref{E_ex}).

To keep the dimensions manageable, we include only neutron excitation
in the $Q$-block spaces,  Beyond that, we only allow seniority-zero
configurations in the neutron spectrum.  The occupation numbers are thus
the same for both orbitals of a Kramers' pair.
We also restrict the dimension of the
\def\Emax{E_{\rm max}}
space keeping only configurations below an energy $\Emax$,
\be
\Eex(q\mu)  \leq \Emax.
\ee  
Here and in the
construction of the full Hamiltonian 
in Sec. \ref{full} 
below
we set $\Emax = 4 $ MeV.

Table \ref{N_K}  presents some characteristics
of the $Q$-blocks constructed in this way.  
The largest block has a dimension
$N_q = 153$  
and total dimension of the bridge configurations
is 514.  These are small enough for  
calculations on laptop computers.  Notice that the 
largest dimensions are in the middle region of the barrier.  
This is consistent with the common understanding that the single-particle
density of states at the Fermi level is higher on top
of the barrier than elsewhere.  
\begin{table}[htb] 
\begin{center} 
\begin{tabular}{|c|cc|cc|} 
\hline 
$Q$ (b) & $N_k$ & $N_p$ & $N_p^{od}$  &
${N_{db}}^{od}$\\
\hline
18 & 42 & 253 & 416& 17\\
22 & 97 & 718 &  1183  & 40\\
26 & 153 & 1391 &  1930 & 77\\
29 & 125 & 1046 &  1109  & 48\\
33 & 65& 434 &  322 & 16\\
37 & 32 & 159 &  & \\
\hline
sum & 514 & &&\\
\hline 
\end{tabular} 
\caption{
Dimension $N_k$ of Q-blocks on the fission barrier of $^{236}$U based on 
the \unedf1 energy functional and an excitation energy cutoff
$E_{\rm cut}=4$ MeV.  The column
$N_{\rm p}$  shows the number of upper off-diagonal pairing matrix elements
in each $Q$-block. Column 4   shows the
number of pairing matrix elements between one $Q$-block and
the next.  Similarly column 5  shows the number of 
diabatic matrix elements.  
} 
\end{center} 
\label{N_K}
\end{table} 
\begin{figure}
\includegraphics[width=1.0\columnwidth]{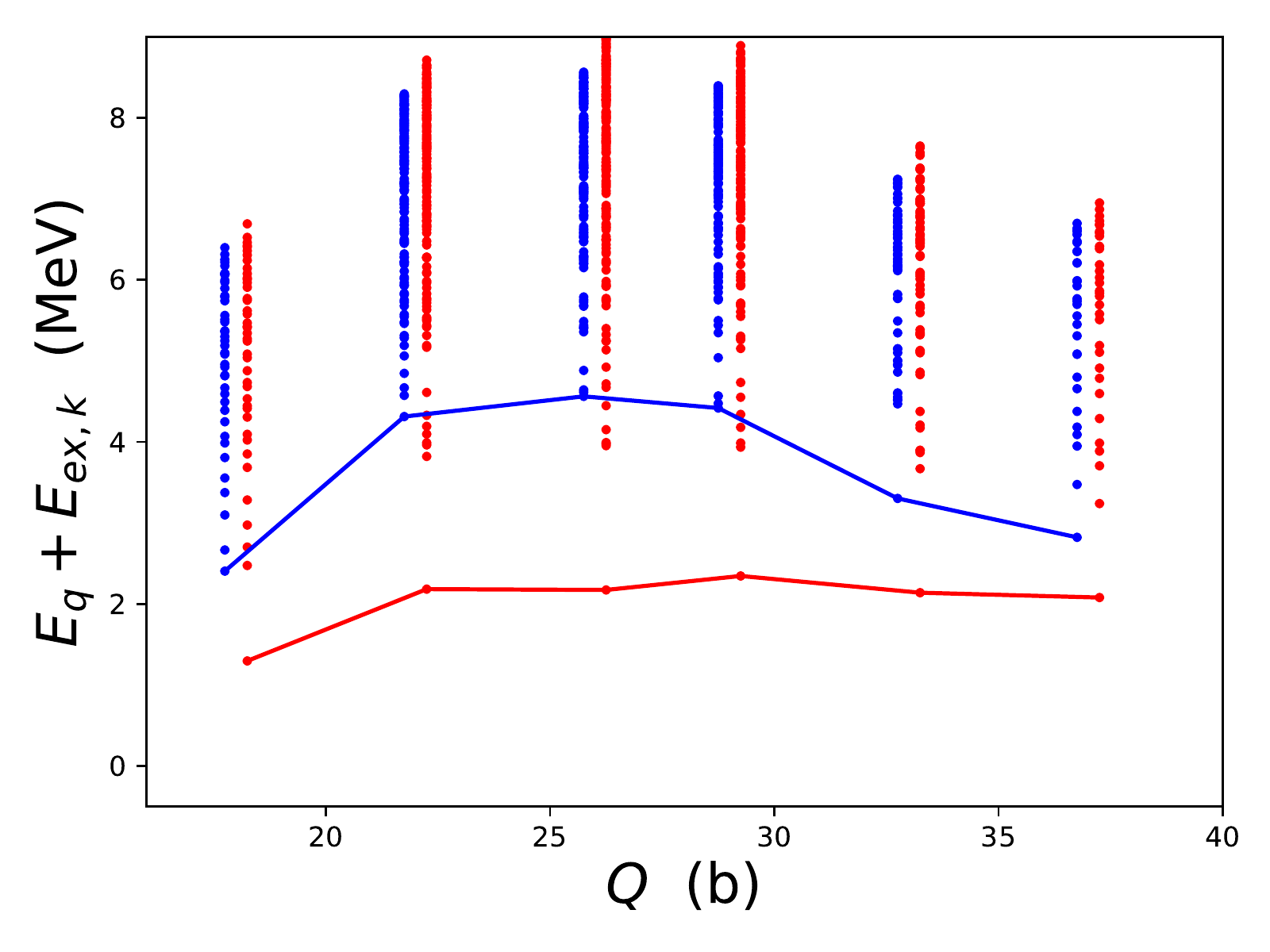}
\caption{
\label{hbridge_diagonal}
Energies of configurations in the Q6  model of $H_{\rm bridge}$.  
The baseline has been shifted by 1 MeV to take into
account the pairing energy in the Q=14 Q-block.
Blue
line: the scaled PES $E_q$; blue dots: diagonal configuration energies
$\langle q\, \mu | H_q | q \,\mu\rangle$
including $V_q$ and the excitation energies $E_{\rm ex}$ of the
excited particle-hole configurations; 
red dots: $Q$-block eigenenergies with pairing interaction included
in $H_q$;  red line: scaled PES with pairing. The cut-off energy of the particle-hole
excitation spectrum is $E_{\rm max} = 4$ MeV and the pairing strength in
the $Q$-block Hamiltonians is $G= 0.2$ MeV.
}
\end{figure}
The diagonal spectrum of $\Hb$ is shown in Fig. \ref{hbridge_diagonal}.
We use these matrix elements in the full Hamiltonian model treated in Sec. \ref{full}  below.

\subsubsection{Interactions}
\label{vp}
  Except for the very lightest systems, microscopic Hamiltonians
rely on a reduction of the interaction terms to an 
effective  two-body nucleon-nucleon interaction, see e.g. Ref.
\cite{ro12}.  In this work, we will use simplified interactions
whose overall strengths are guided by previous experience.  
There are two kinds of 
interaction that can mix configurations in the
seniority-zero configuration space.  The first is 
the pairing interaction, which is crucial for promoting
spontaneous fission \cite{ro14}. It is implicit
in the BCS and HFB approximations, but must be explicitly
included as a residual interaction in a HF-based configuration
space.  Following common practice, we parameterize it as the Fock-space operator 
\be
\hat v_{\rm pairing} = -G_{qq'} \sum_{i \ne j} a^\dagger_{i} a^\dagger_{\bar i}
a_{\bar j } a_j. 
\label{v_p}
\ee
Here $i$ and $\bar i$ are time-reversed partner orbitals.

\begin{figure}
\includegraphics[width=1.0\columnwidth]{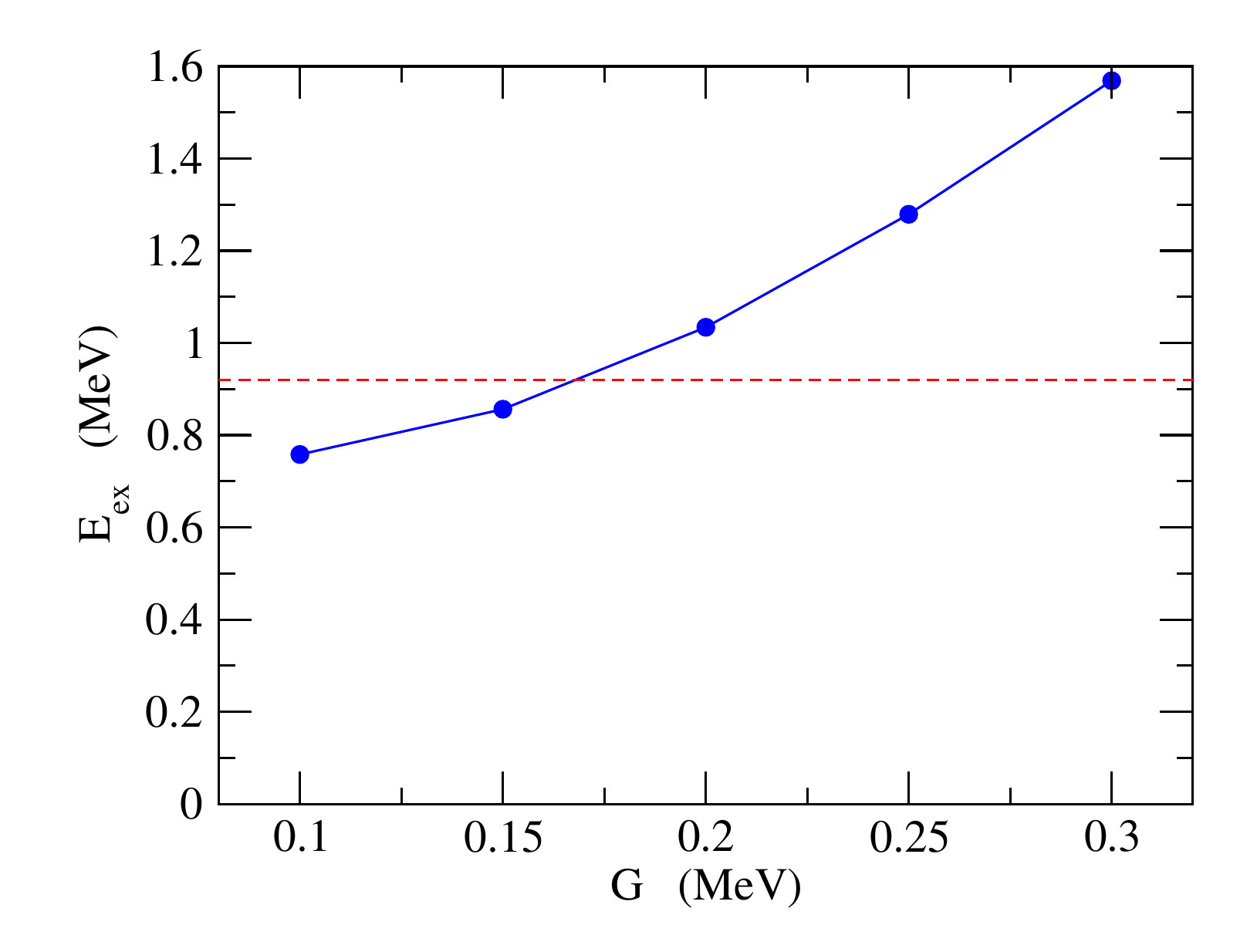}
\caption{
\label{G-fig}
Excitation energy of the first $K^\pi = 0^+$ excited state
in the spectrum generated from the ground-state reference
state at $Q=14 $ b as a function of the pairing strength $G$.
Other parameters are the same as used in $\Hb$.
}
\end{figure}

We next determine the interaction strength $G=G_{qq}$ within the
$Q$-blocks. The effective strength depends on the size of
the configuration space; see Ref. \cite{pi05} for numerical studies
of that dependence.
A typical BCS calculation might be carried
out in a full major shell; the observables such as the odd-even
binding energy differences can be fitted  with a pairing strength 
$ G \approx 25/A $ MeV.  This  gives $G \approx 0.1 $ MeV in
the actinide region.  However, for our much more limited space
the strength should be larger.  We choose to set the strength
to reproduce the excitation of the first excited $0^+$ state in
the seniority-zero configuration space of \uf, $E_{\rm ex}(1) = 0.92$ MeV.  
This yields $G \approx 0.17$ MeV as may be seen in Fig. \ref{G-fig}.
This is close to the value $G = 0.2$ MeV that we use within the $Q$-blocks 
in the full Hamiltonian.  

The pairing strength has to be modified
for matrix elements between configurations in different $Q$-blocks.
The  general formula \cite{lowdin} 
for calculating two-body matrix elements in a nonorthogonal CI basis 
could be used, but it is very time-consuming to carry out.  Another
formula based on the generalized Wick's theorem \cite{ba69} is fast.
However, it requires the two configurations to have a nonzero overlap
which is hardly the case for the pairing interaction.  In our
present model, we will simply assume that overlaps of the configurations
attenuate all matrix elements by the same factor,
\be
G_{qq'} = c\,  G.
\ee
Here $q$ and $q'$ are neighboring Q-blocks and
$c = e^{-1}$ is a constant set by the target overlap distance
$\Delta \zeta = 1$.

The second kind of interaction matrix element is the coupling to
diabatic partner configurations.
The diabatic matrix elements are nonzero only for configurations that have
large overlaps, so the 
generalized Wick's theorem can be applied to calculate them.
However, we would
still like to make simplifying approximations that make the
model calculations more transparent.
A convenient functional form for parameterizing the interaction is  \cite{388}
\be
\langle \Psi_{q\mu} | \hat v_{db} | \Psi_{q'\mu}\rangle
\label{v_db}
=\ee 
$$
\langle q\mu| q'\mu\rangle 
\left(\frac{1}{2}(E_{q\mu} + E_{q'\mu}) - 
h_2(\bar Q) (\Delta \zeta)^2\right)
$$
where $\bar Q = (q + q')/2$ 
and
$E_{q\mu}$ is the energy of
the configuration including the modified PES. For the present study
we will assume a fixed value for the interaction strength, $h_2=
1.5 $ MeV. The motivation  for the functional form of Eq. (\ref{v_db})
and the choice of the strength parameter $h_2$ are discussed in Appendix B.  
The formula is implemented in the Q6 model with 
$\langle q\mu| q'\mu\rangle=e^{-1}$ 
for neighboring Q-blocks and $\langle q\mu| q'\mu\rangle=0$ 
when $q$ and $q'$ are farther away from each other.
 
\begin{figure}
\includegraphics[width=1.0\columnwidth]{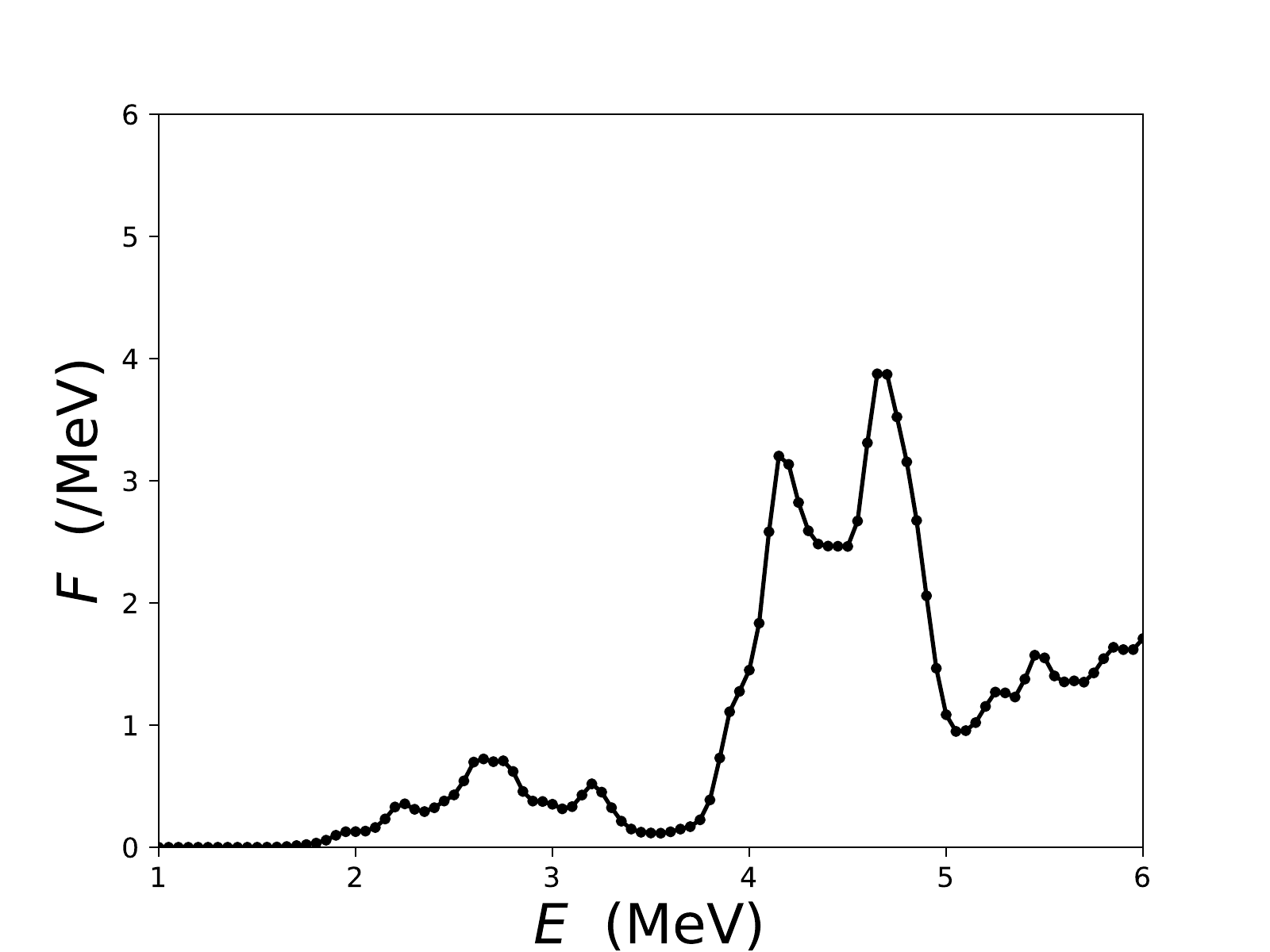}
\caption{\label{Hb_inv}
Transmittance of $\Hb$ as measured by Eq. (\ref{Hbinv-eq}).
The smearing parameter $w$ is set to be 0.2 MeV.
}
\end{figure}

Before going on to the full Hamiltonian we can get a sense
of the transmittance of $\Hb$ by calculating the quantity
\be
F(E) = \sum_{\mu \in q_a} \sum_{\mu' \in q_b} |(\Hb -
S(E-iw))^{-1})_{\mu\mu'}|^2.
\label{Hbinv-eq}
\ee
Here  $q_a$ and $q_b$ are the first and last $Q$-blocks in $\Hb$ and
$w= 0.2 $ MeV is an averaging parameter. The 
result is shown in Fig. \ref{Hb_inv}.  There is a large peak just
above 4 MeV which is just below the barrier top at 4.6 MeV.
There are also smaller peaks below 3.5 MeV that are probably
composed of strongly paired configurations.  The lowest ones can 
be identified with the eigenvalues of the $\Hb$ spectrum.  
\subsection{The full Hamiltonian}
\label{full}

It remains to add the two GOE reservoirs to complete
the Hamiltonian depicted in Fig.1. As discussed in Appendix \ref{appB}
we have a certain freedom to set the dimension of a GOE reservoir provided
the decay widths are modified to keep  the transmission factors
 Eq.~(\ref{NTF}) fixed. The relevant properties of the entrance  and capture channels 
are well-known experimentally,  and we set the transmission coefficients
accordingly.  Somewhat arbitrarily,
we set the dimension of the reservoirs to $N_{\rm GOE} = 100$
and the internal interaction strengths in the GOE Hamiltonian to 
$ v = 0.1$ MeV. This produces a level density
of $\rho_0 = 31.8$ MeV$^{-1}$ in the middle of the spectrum.  With 
$\Gamma_{\rm in}=10$ keV, the 
resulting transmission factor for an $s$-wave neutron entrance channel
at $E_n = 1$ keV is $T_{\rm in}=0.02$.  
The scaled capture width of the GOE states
is $\Gamma_c = 1.25$ keV.
As discussed elsewhere, the fission reaction probability $\Ts_{{\rm in},f}$ is rather 
insensitive to the partial widths in reservoir $b$; we have 
chosen the value $\Gamma_f = 15 $ keV.  This is well within the plateau region.

\begin{figure}
\includegraphics[width=1.0\columnwidth]{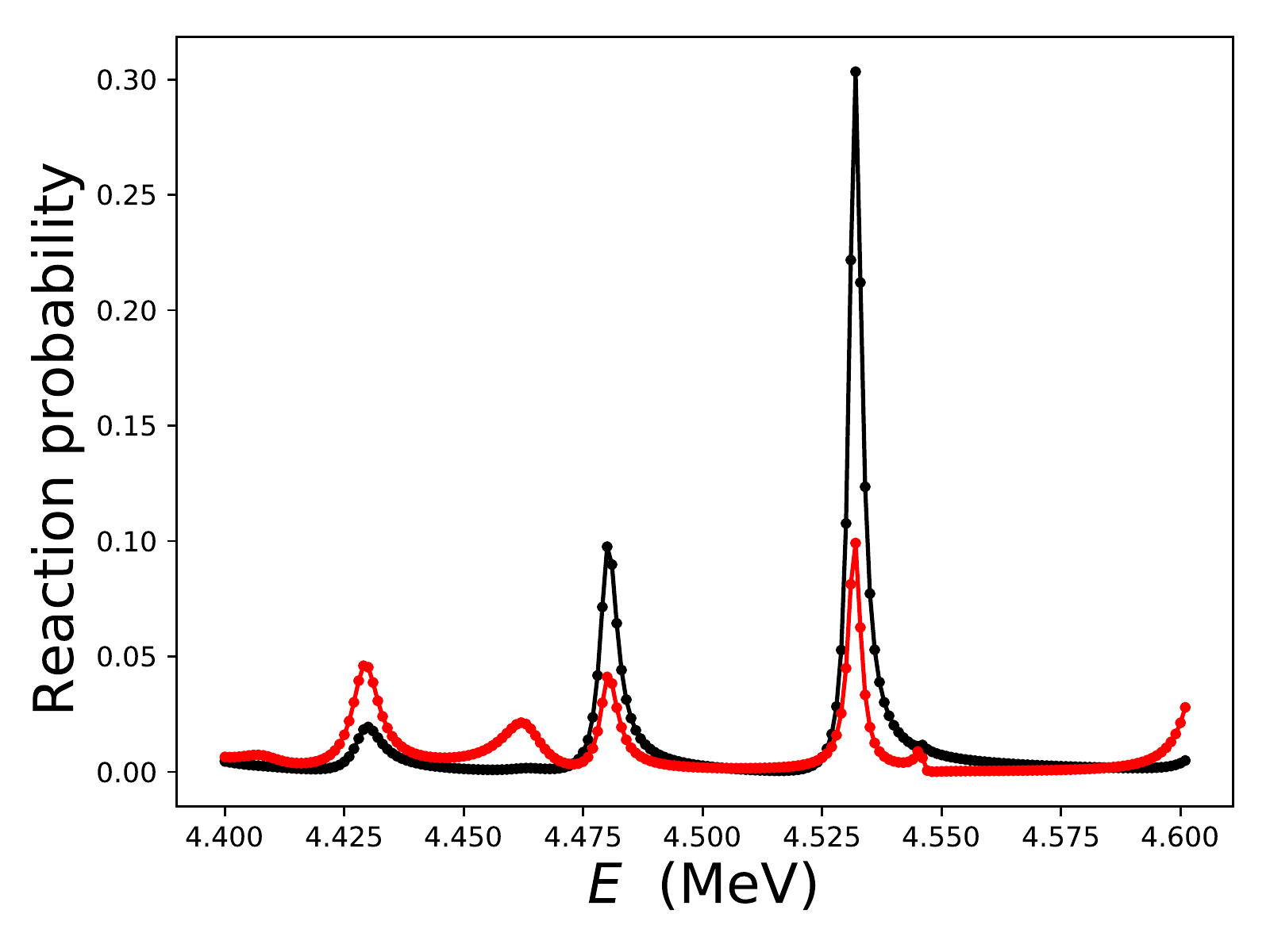}
\caption{
\label{resonances}
Resonances of the full Hamiltonian at $V_b = 4$ and $E$ 
around 3.5 MeV.  The black and red points show reaction probabilities
for $T_{\rm in,c}$ and $T_{\rm in,f}$ respectively.
}
\end{figure}

Two sets of interaction matrix elements are still needed to 
have a complete Hamiltonian, namely those between the GOE reservoirs 
and $\Hb$.  These are placed as depicted in Fig. 1.  We parameterize these as Gaussian-distributed random
variables with rms strengths $v_a$ and $v_b$.
Each set connects all of the states in the reservoir
to all of the configurations in the  adjacent $Q$-block.  Unfortunately, 
the strength of these interactions cannot be calculated from microscopic
nucleon-nucleon Hamiltonians 
without a better understanding
of the structure of the reservoir states.  Thus the overall
magnitude  of the fission branch is beyond the scope of the 
model.  Nevertheless, the model can still shed light on aspects
of the barrier-top dynamics.  One aspect is the energy dependence of the 
reaction probabilities, and another is the importance of 
the diabatic interaction in the bridge dynamics.  These are
discussed in the next 
section.  
For a baseline model we take
$ v_a = 0.02$ and $v_b = 0.03$ MeV.  With these parameters the
branching ratio $\alpha^{-1}$ can approach the order of magnitude
seen experimentally.  Figure \ref{resonances} shows the
reaction probabilities for the Hamiltonian in a small
interval of energy.  The entrance transmission factor is
small enough to show individual compound-nucleus resonances.

\section{Reaction probabilities}

\subsection{Energy dependence}

\begin{figure}
\includegraphics[width=1.0\columnwidth]{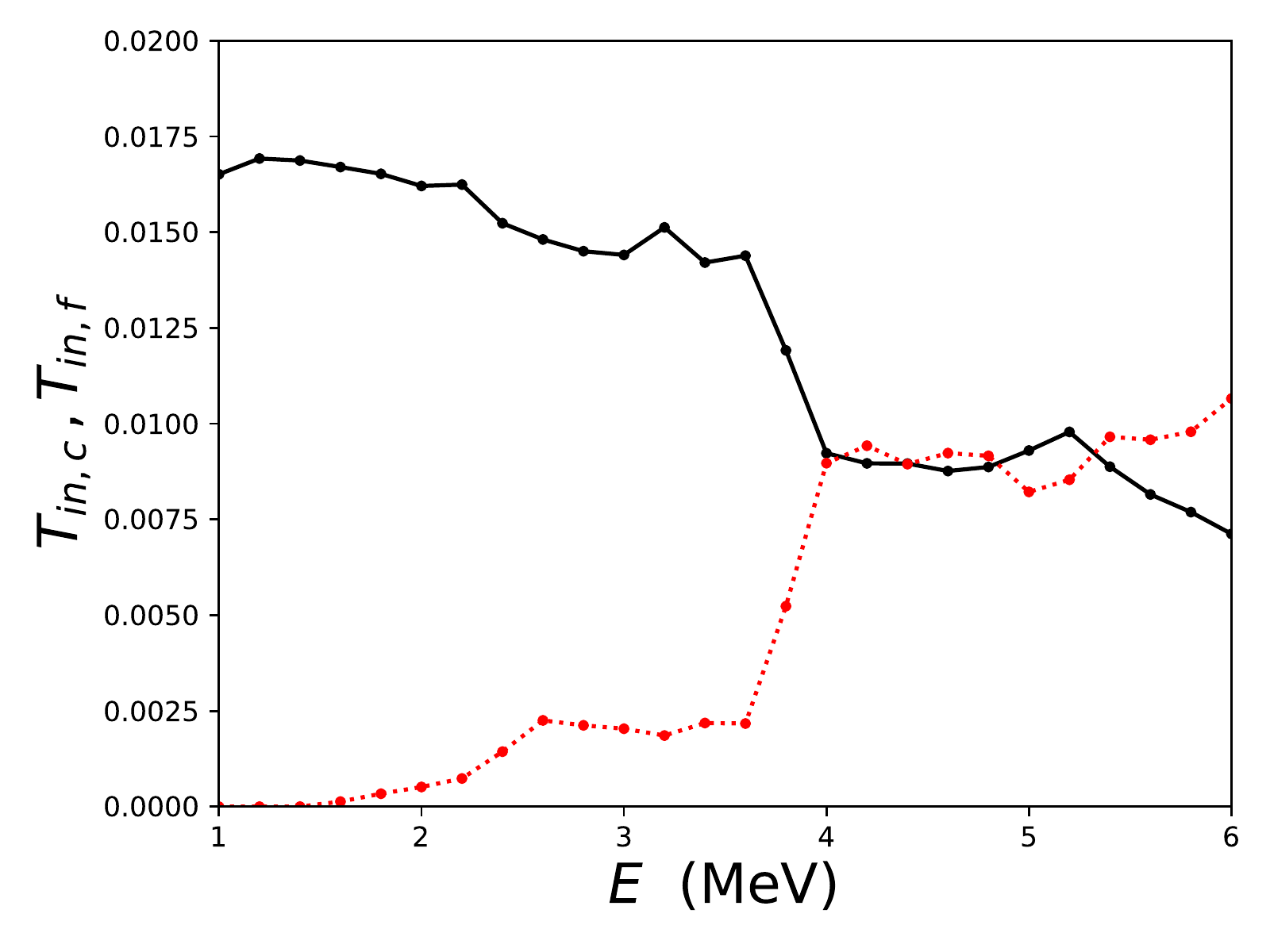}
\includegraphics[width=1.0\columnwidth]{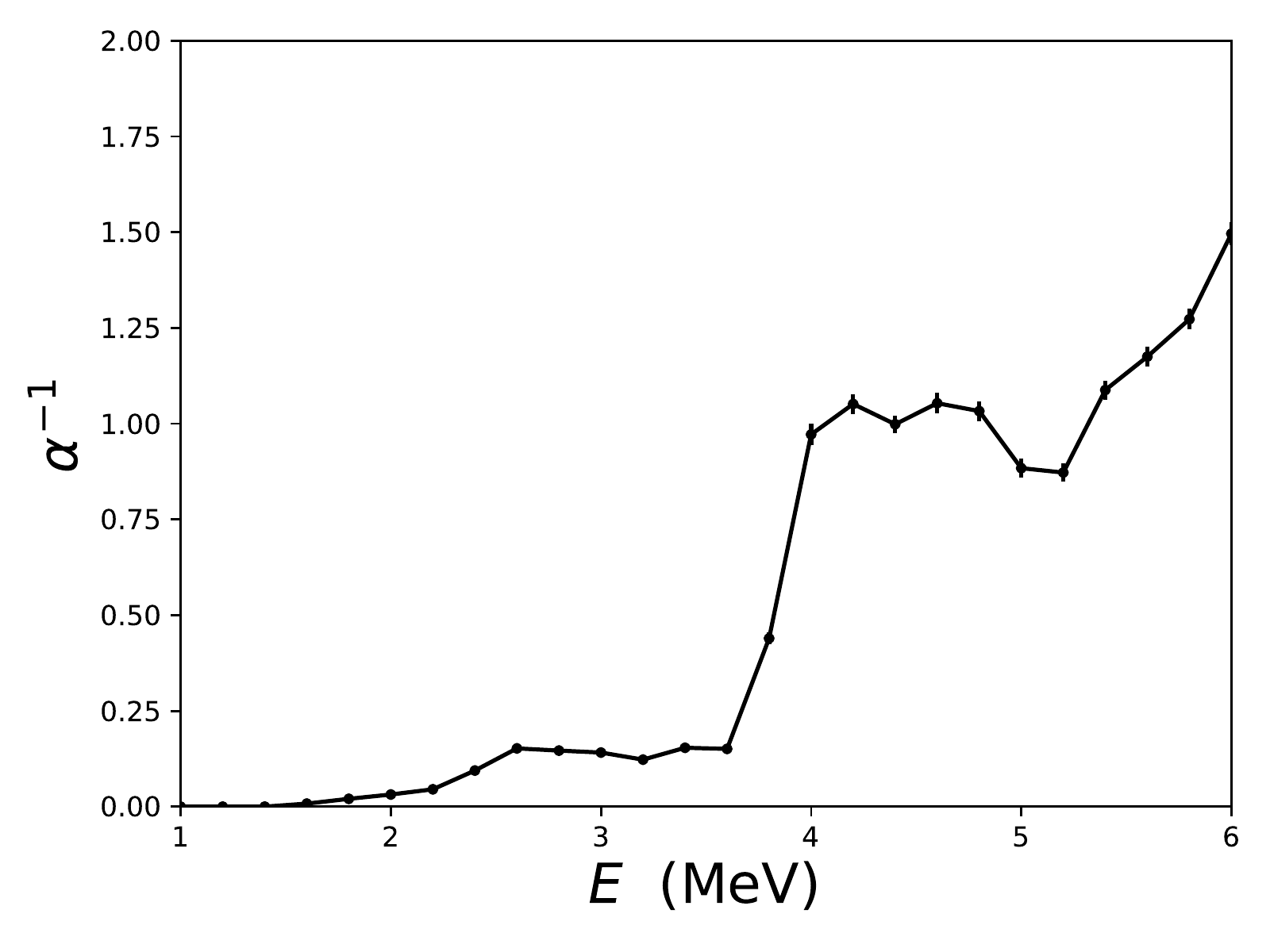}
\caption{
Average reaction probabilities for capture (black circles)  and
fission (red circles) as a function of energy
$E$ are displayed in the upper panel.  The branching ration
$\alpha^{-1}$ is shown in the lower panel. 
In the calculation, the two GOE reservoir Hamiltonians are
centered at $E$ with fixed decay widths to isolate the energy dependence
of transmittance through $\Hb$. 
\label{TvsE}
}
\end{figure}

In this work, we are mainly interested in average reaction probabilities . 
The averages are
obtained by integrating over some interval of energy that includes
multiple resonances, and then averaging over the random
GOE samples in the Hamiltonian.  Fig. \ref{TvsE} shows
the reaction  probabilities for  capture and fission
calculated this way. The points were obtained by integrating
over an interval of 0.5 MeV and averaging over 400 GOE samples.
%
Notice that the total reaction probability remains fairly
constant at $\Ts = \Ts_{{\rm in},c} + \Ts_{{\rm in},f}  \approx 0.02
$ over the entire range plotted.
This is required of compound nucleus theory when the
entrance channel transmission factor is small compared 
to the others.  Notice also that the fission probability
does not increase smoothly at subbarrier energies.  This
goes against the Hill-Wheeler barrier-penetration formula.
There are small windows well below the barrier 
for transmission that are probably due to the paired $Q$-block 
ground states. Note also that the reaction
probability for fission is monotonically increasing in the energy
region shown, contrary to Eq. (\ref{Hbinv-eq}).

\subsection{Branching ratio}

The branching ratio $\alpha^{-1}$ (Eq. (\ref{alpha})) as a function
of energy $E$ is displayed in the bottom panel of Fig. \ref{TvsE}
as a function of energy $E$.  The ratio roughly tracks the same irregular increase
as that found in the reaction probability shown in the upper panel. 
It reaches
a level of $\alpha^{-1}\approx 1$ at the higher energies. 
This  is less than the experimental ratio
of $\approx 3$ in the fission of \ut~by low-energy neutrons.  
Since the experimental order of magnitude is achieved, the model should
be useful for qualitative insights into the transport mechanisms.

We next examine the dependence of the branching ratio on 
the Hamiltonian parameters.    Table \ref{parameter_variation}
presents the results of calculations with different sets of 
parameters.
The calculation for a baseline set of parameters is shown
on the top line of the table.
\begin{table}[htb] 
\begin{center} 
\begin{tabular}{|c|cc|cccc|c|} 
\hline 
Model & $\Gamma_c$ & $\Gamma_f$ & $v_p$ & $h_2$ & $v_a$ & $v_b$ 
&   $\alpha^{-1} $ \\ 
\hline
  base  & 0.00125 & 0.015       & 0.2 & 1.5  & 0.02  & 0.03 & $0.95$\\
   A & 0.0025 &                 &     &      &       &      & $0.55$  \\
   B &        &   0.03          &     &      &       &      & $1.14$  \\
   C &        &   0.045         &     &      &       &      & $1.23$ \\
   D &        & 0.15            &     &      &       &      & $1.65$ \\  
   E &        &                 &     &  3.0 &       &      & $1.10$ \\
   F &        &                 &     &  0.0 &       &      & $0.13$\\  
   G &        &                 &  0.1&      &       &      & $ 0.37$  \\
   H &        &                 &     &      & 0.01  &      &  $0.59$  \\
   I &        &                 &     &      & 0.04  &      &  $1.29$ \\
   J &        &                 &     &      &       & 0.015& $0.60$ \\
   K &        &                 &     &      &       & 0.06 & $1.20$ \\
\hline 
\end{tabular} 
\caption{
Branching ratios calculated with Eqs. (\ref{alpha}) and
(\ref{datta}) for several sets of 
energy parameters. Units are
MeV.
The base parameters are given in the top line.  For the other cases
only the changes from base are shown in the table.   
In the calculations, Eq. (\ref{alpha}) was evaluated 
by averaging over an interval from 4.25 to 4.75  MeV.  The
column shows the mean branching ratio obtained with 400
samples of the compound-nucleus GOE.  The resulting in  uncertainty limits
are  about $\pm 0.02$.
\label{parameter_variation}
} 
\end{center} 
\end{table} 
It gives  $\alpha^{-1}\approx 1.00 \pm 0.02 $ at 4.5 MeV which
is still well below the observed value $\alpha^{-1}\approx 3$ at
the physical neutron threshold at 6.5 MeV.
One obvious reason is that the excited states 
of the protons 
have been left out.  Their  inclusion might
increase the branching ratio.  Also, the off-diagonal
neutron-proton matrix elements are not active due to the zero-seniority
structure of the configurations.  However, if
no reasonable parameter sets can be found to reproduce the experimental
$\alpha^{-1}$ in seniority-zero configuration space,
it would be indirect evidence that the 
space must be extended to include the far more numerous broken-pair
configurations.

The other entries in the table indicate the sensitivity of $\alpha^{-1}$
to the Hamiltonian parameters.  Lines A-D show the dependence on the decay 
widths, $\Gamma_i$.
Entry A is a preliminary check on the model to confirm that an increase in the capture branch
produces a corresponding decrease in the fission branch.  This
is expected in compound nucleus theory when there are many channels
for each branch.  We see from the B to D entries that the branching
ratio is insensitive to the fission branch over a wide range of fission
widths.
The entries E -K  test the dependence on interaction parameters
in the Hamiltonian.  Entries E and F show that the diabatic interaction cannot
be ignored, but the ratio is insensitive
to increase beyond the value in the baseline Hamiltonian.
One sees from entry G that an  error in the pairing strength
is likely to propagate to a similar relative error in the branching ratio.
This may be contrasted with spontaneous fission, 
where theoretical lifetimes are very strongly dependent on the pairing
strength \cite{ro14}.  
Entries H - K  in the  table show the
effect of changing  the  matrix elements between $\Hb$ and the
GOE reservoirs.  As expected, weaker  interactions produce
smaller fission probability.  Doubling $v_b$ from  its baseline
value does not make a significant change in $\alpha^{-1}$, as might
be expected from the experience with the fission decay widths.
However, there is 
a substantial decrease when $v_b$ is reduced, indicating
that the baseline value is at the beginning of a plateau.

\section{Conclusion and Outlook}

The model Hamiltonian in this work introduces for the first
time 
a CI computational framework 
to describe the many-body dynamics at the fission barrier. 
A primary conclusion of the study is that the transport
appears not to be carried by a small number of internal
channels, but rather is diffuse and spread over many
barrier topic configurations.  If so, it invalidates the
transition-state theory that has been accepted uncritically 
since the earliest work on the subject.  However, the model
may be deficient in a way  that could alter that
conclusion.  For example, the  space of wave functions was generated
with time-even  constraints which produce only time-even paired wave functions.
These have limited band width to transport flux, as was
demonstrated in Ref. \cite{be22}.  If one added time-odd
configurations by constraining with a collective
momentum operator \cite{hizawa2021,hizawa2022} 
as well, the bandwidths would certainly
increase.

On the other hand, increasing the space and the scope of
the Hamiltonian in other ways is not likely to bring
the model closer to the transition-state physics.  The
pairing interaction acts independently in the neutron
and proton subspaces, so inclusion of seniority-zero
proton configurations would not make a qualitative 
change in the excitation function.

The off-diagonal proton-neutron interaction matrix elements 
may become dominant when broken-pair configurations are
included in the CI space\cite{bush92}, and they may work against
the collectivity promoted by the pairing interaction.
In the limit of large off-diagonal elements with
random signs, the dynamics would become diffusive.
This probably happens anyway at large excitation
energy, but the question remains open for barrier-top
energies.

One conclusion points  favorably toward future efforts
to build a microscopic theory of fission.  one sees that
the transport properties are determined around the
barrier as in the transition-state theory.  The branching
ratios can thus be calculated
without detailed information about the post-barrier
Hamiltonian.  We called this the ``insensitivity property".
The qualitative explanation is very simple:  once
the system gets past the barrier, it can go 
so many directions in phase space to get to a fission 
channel that one can neglect the possibility
that it may come back.

We also investigated the relative importance of pairing
and diabatic interactions.
As expected, the branch ratio is quite sensitive  to the pairing 
interaction strength.  
In fact the nucleus would
not fission at
barrier-top energies without pairing being included in some
way in the GCM or time-dependent HF approximation \cite{scamps2015,tanimura2015}.
In contrast, the diabatic interaction is not
essential for fission, but it  substantially enhances
the fission branch at a physically relevant strength level.

The prospects for making the model more realistic
depend very much on the size of the configuration
space in $\Hb$.  Some dimensions for extended spaces
are shown in Table \ref{SpaceExtension}.  
The costliest numerical task in the reaction theory is
the matrix inversion in Eq. (\ref{datta}), but
it can be speeded up by taking advantage of its 
tridiagonal block structure \cite{ha20,pe08}.
Inclusion of proton excitations in
the zero-seniority model space requires only $Q$-block dimensions
of the order of a few thousands. This is certainly feasible,
even with the limited computational power of desk-top
computers.  
\begin{table}[htb] 
\begin{center} 
\begin{tabular}{|c|ccc|cc|} 
\hline 
& \multicolumn{3}{|c|}{seniority zero} &\multicolumn{2}{|c|}{all $K^\pi= 0^+$}  \\
\hline
$q$ & $n$ only & $p$ only  &  $n+p$ & $n$ only  & $n+p$ \\
\hline
 18   & 42  & 23   & 966     & 738  & $3.2\times 10^4$  \\ 
22    & 97 & 46   & 4462    & 3088 & $3.5\times 10^5$  \\
26 & 153 &  25   & 3825    & 8232 & $3.1\times 10^5$   \\
29 & 125 &   33   & 4125    & 5080 & $4.3\times 10^5$   \\ 
33 & 65 &   18   & 1170    & 1455 & $1.7\times 10^4$ \\ 
37 & 32  &  43    & 1419    &  409   & $3.9\times 10^4$  \\
sum & 514 & 188 & 15967 & 19002    & $1.1\times 10^6$   \\
\hline 
\end{tabular} 
\caption{ 
Dimensions of extended spaces to include proton configurations
and all seniorities. The cutoff in the configuration spaces
is $E_{\rm max} = 4$ for both neutrons and protons. The seniority-zero 
all-nucleon space thus 
extends up to 8 MeV.
\label{SpaceExtension}
} 
\end{center} 
\end{table} 

Including all seniorities
in the $Q$-block configuration space is much more challenging.  The last column
in Table III shows the resulting dimensions. 
The number of $K^\pi = 0^+$ configurations with $\Emax = 4$  MeV
is of the order $10^{5}$. With 6 reference states in the bridge
region the total dimension is $10^6$.  
To put this in perspective, shell model diagonalizations
have been reported for configuration-space
dimensions of the order of 10$^{10-11}$ \cite{shimizu2019}.  

Instead of taking brute-force approach to the large configuration
spaces, it might be more productive to look for more 
sophisticated schemes to truncate the active space of states.
The theory is already a statistical one due to the
GOE reservoirs, but we have not been able to avoid the time-consuming
task of numerically
sampling the GOE Hamiltonians.
Eq. (\ref{Hbinv-eq}) 
was an attempt to estimate the transmittance of $\Hb$ without
the Monte Carlo sampling, but the accord with the full Hamiltonian 
is not satisfactory.
Finally, we need a better understanding of statistical aspects
of the interaction matrix elements, since calculating them
individually is out of the question.

So far the model does not provide a crisp answer to the question,
``How many channels are active in barrier-top fission"?  There are
at least two ways that one could investigate the question.  One is to examine
how  the probability 
flux between $Q$-blocks 
is distributed
over the linkages between the block eigenstates: many active links
imply many channels.   Another way is to examine the  resonance
width fluctuations in the region of isolated resonances.  
 The fission widths should satisfy the formula\cite{va73}
\be
\nu = \frac{2 \langle \Gamma_f \rangle^2}{\langle \Gamma_f^2 \rangle - \langle \Gamma_f \rangle^2} 
\ee
where $\nu$ is the effective number of channels.  
We intend to investigate this issue in a future publication.

The main codes used in the work will be available on request and 
later in the Supplementary Material accompanying the published article. 

\begin{acknowledgments}
We thank L.M. Robledo for providing the HFBaxial codes, 
and G. Col\'o for pointing our attention to Refs. \cite{yo86,co94}. 
We also thank J. Maruhn and P-G. Reinhard for helping in adapting 
the Skyax code to provide needed energies and orbital properties
for constructing $\Hb$. 
This work was supported in part by JSPS KAKENHI Grants No. JP19K03861 
and No. JP21H00120.
\end{acknowledgments}

\appendix

\section{GOE model of the statistical reservoirs}
\label{appB}
This appendix reviews the basic properties of the GOE as a
model for the compound nucleus and other statistical reservoirs.
It is characterized by two parameters, 
the dimension of the space $N_{\rm GOE}$
and the strength of the Gaussian-distributed residual interaction
$\langle v^2\rangle^{1/2}$.  We shall also refer to the level
density at the center
of the distribution, given by 
\be
\rho_0 = N_{\rm GOE}^{1/2}/\pi \langle v^2\rangle^{1/2}.
\ee 
For a model of 
a compound nucleus that can decay by gamma emission (capture) or
by fission, there are five additional parameters.  
They are the decay widths $\Gamma_{\rm in},\Gamma_{\rm cap}$
and $\Gamma_{\rm f}$, together with the number\footnote{The entrance
channel is unique, i.e. $N_{\rm in} = 1$.} of capture and
fission channels, $N_{\rm cap}$ and $N_{\rm f}$.  Each channel is 
paired with a state $\mu$ in the GOE space and given the appropriate
decay width $ E_\mu \rightarrow E_\mu - i\Gamma/2$.
In compound-nucleus phenomenology the couplings between the
channels and the reservoir are better parameterized by transmission
coefficients defined as
\be
T_{i} \equiv 2 \pi \rho_0 \Gamma_{i}.
\label{NTF}
\ee

We will now set the GOE parameters for the compound-nucleus 
treatment of the $n+^{235}$U $\rightarrow$ $^{236}$U$^*$ reaction.
The transmission factor for the  entrance channel is taken  from the optical model
systematics; it is roughly parameterized as
\be
T_{\rm in} = 2 \pi S_0 E_n^{1/2}
\ee
where $S_0 \approx 10^{-4}$ is the strength function \cite[ Fig. 10]{ripl3} 
and $E_n$ is the 
neutron bombarding energy in eV units.  For our numerical studies below 
we take $E_n = 10$ keV which implies $T_{\rm in} = 0.063$.  
The average gamma decay width of the states in the reservoir
is $\Gamma_{\rm cap} \approx 0.04$ eV.  The empirical level density
associated with an entrance channel is $\rho_0 \approx 1$ eV$^{-1}$, giving
\be
T_\gamma = 2 \pi \rho_0 \Gamma_{\rm cap} \approx 0.25. 
\ee
We also know that
there are many gamma decay channels, so $N_{\rm cap} \gg 1 $.
   
It is not as easy to specify the coupling to the fission channels.
For the moment 
we take the fission width to be $\Gamma_{\rm f} = 0.42$ eV 
as in an example from Ref. \cite{BK17}.
From the empirical data one can only extract qualitative information
about the number of exit channels.  
As a simple exercise to see how the physical observables depend on
the GOE parameters, we take the above parameters plus 
$(N_{\rm GOE},N_{\rm cap},N_{\rm f}) = (50,10,1)$ as a baseline
for numerical modeling.  

A key attribute of the compound nucleus is that its decay properties
are independent of how it was formed, subject to some well-known caveats.  The
independence  is
encapsulated in the compound nucleus formula for $\Ts$:
\be
\Ts_{\rm in,i} = T_{\rm in}\frac{T_i}{\sum_{i} T_{i}}. 
\label{CN}
\ee
If the entrance channel width is small compared to other decay
widths,  the reaction probabilities should sum to $T_{\inn}$:
\be
 T_{\rm in} \approx \sum_i \Ts_{\rm in,i}.
\label{T=T}
\ee
Table \ref{GOEexamples} shows how well this works for
several treatments of the dimensions
$N_{\rm GOE},N_{\rm cap}$ and $N_{\rm f}$.  
\begin{table}[htb] 
\begin{center} 
\begin{tabular}{|c|ccc|cc|} 
\hline 
Model & $N_{\rm GOE}$ & $N_{\rm cap}$ & $N_{\rm f}$ & $\sum_i T_{\rm in,i}$ &  $\alpha^{-1}$ \\ 
\hline
  A & 50 & 10 & 1& 0.051 &  2.03  \\ 
  B & 100 & 10 & 1& 0.050 &  1.98 \\
  C & 800 & 10 & 1& 0.047 &   2.07 \\
  D & 50 & 20  & 1& 0.050 &  1.99  \\ 
  E & 200 & 20 & 1&  0.049 & 2.00 \\ 
  F & 50 & 10 & 2&  0.054 &  3.75 \\
  G  & 50 & 10 & 10& 0.057&  8.57 \\ 
\hline 
\end{tabular} 
\caption{ 
Reaction probability and branching ratio Eq. (\ref{alpha}) for
GOE models of the $n+^{235}$U compound nucleus reactions.  The
nominal transmission factors are $(T_{\rm in}, T_{\rm cap}, T_{\rm f}) =
(0.063,0.25, 2.64)$. The integration interval for calculating
the branching ratio covers the center third of the GOE
eigenspectrum.  Statistical errors associated with the GOE
sampling are about 1\%.
\label{GOEexamples}
} 
\end{center} 
\end{table} 
One sees that Eq. (\ref{T=T}) is quite well satisfied and
is independent of the dimensional parameters, at least
in the range we have computed.

One of the most important physical observables is the branching
ratio.  
The calculated results for GOE model are shown in the last column of 
Table \ref{GOEexamples}.  The dimensions of the GOE space $N_{\rm GOE}$
is varied in the first three lines, which gives one confidence that
the enormous size of the physical space is not an obstacle to 
constructing a practical model. The branching ratio is also nearly
independent of $N_{\rm cap}$, provided that the number is large.
However, the models F and G show that there is a strong dependence 
on $N_{\rm f}$.  
This is 
a well-known phenomenon and is included in compound-nucleus
theory as the Moldauer correction factor \cite{va73,mo75}.

\section{Insensitivity to fission widths}

\begin{figure}
\includegraphics[width=0.5\columnwidth]{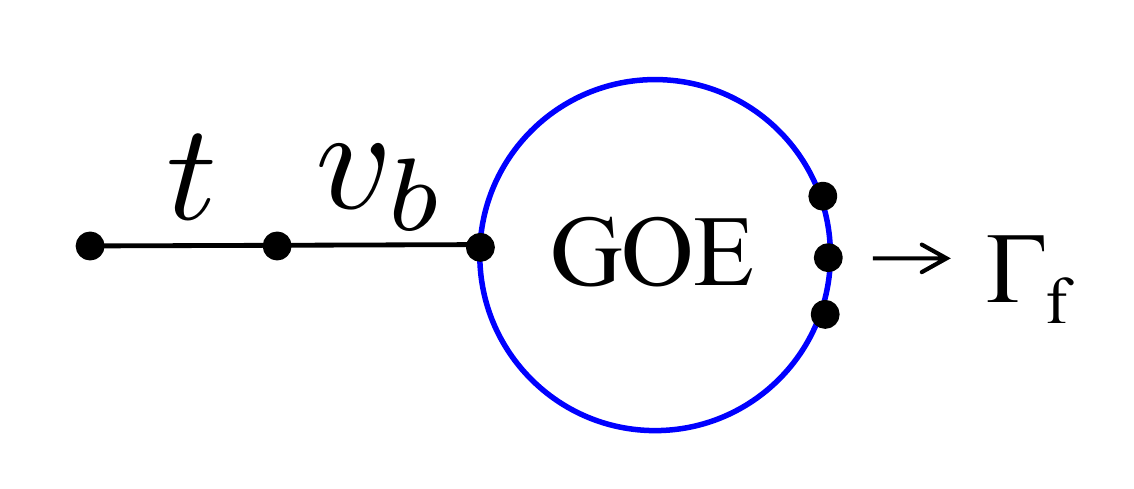}
\caption{
\label{GOE+2}
Hamiltonian structure to test the transmission properties of
the post-barrier reservoir.  The Hamiltonian parameters used
in the calculations for Fig. \ref{insensitivity1} below are:
diagonal energies $E_i = 0$ for the two states $i$ in 
the entrance channel;
$t = 1$ for the interaction linking those states;  $v_b = 0.6$
for the interaction connecting the entrance channel to a state
in the GOE matrix; $\langle v^2\rangle^{-1/2} = 0.1$ for the
internal interaction strength in the GOE matrix; 
and $N_{\rm GOE}=100$ and $N_{\rm f}=100$ for the dimension of the 
GOE Hamiltonian and the number of fission channels.} 
\end{figure}

The fission widths in the model are incorporated
into the GOE of the post-barrier reservoir.  It would
be difficult to calculate those widths from
a microscopic Hamiltonian.  However, we expect that the
dimension of the
post-barrier reservoir is 
largely independent of fission exit channels.
In that situation the
effective decay width is controlled by the coupling to the bridge
states \cite{385}.  
As an example, Fig. \ref{GOE+2} shows the
structure of a simple GOE model to test the sensitivity
to the final-state decay widths.  In it, 
the entrance channel is represented by a chain of two states
that couple to the GOE reservoir.
\begin{figure}
\includegraphics[width=1.0\columnwidth]{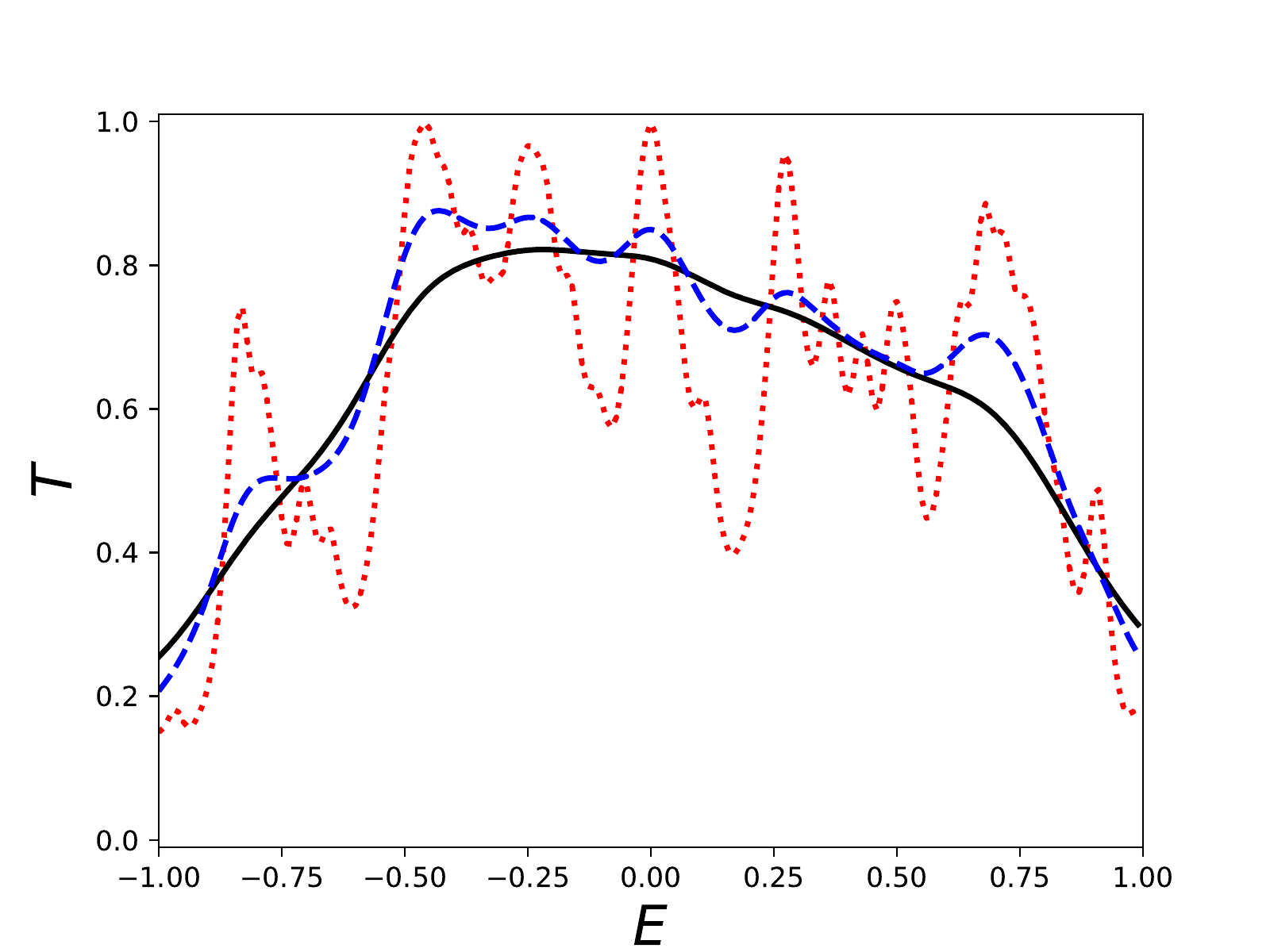}
\caption{
Transmission probability $T_{\rm in,f}$ to decay channels in the post-barrier GOE for
a Hamiltonian with the connectivity shown in the previous figure.
  The decay widths
$\Gamma_f$ of the GOE states are:   0.05 (dotted red line); 
0.2 (dashed blue line);  and 0.4 (solid black line). The other 
Hamiltonian parameters are given in the caption to Fig.
\ref{GOE+2}.
}
\label{insensitivity1}
\end{figure}
In Fig. \ref{insensitivity1}, the reaction probability
$\Ts$ is plotted as a function of energy for a range of final
state widths. One sees that the
average $\Ts$ remains the same over an 8-fold increase in 
$\Gamma_f$.  In this situation, the entrance transmission factor  is approximately
given by
\be
T_{\rm in} = (2\pi \rho_0)^2 \langle v^2 \rangle
\ee
where $\langle v^2 \rangle^{1/2}$ is the average interaction
matrix element between the entry chain and the reservoir.

\begin{figure}
\includegraphics[width=1.0\columnwidth]{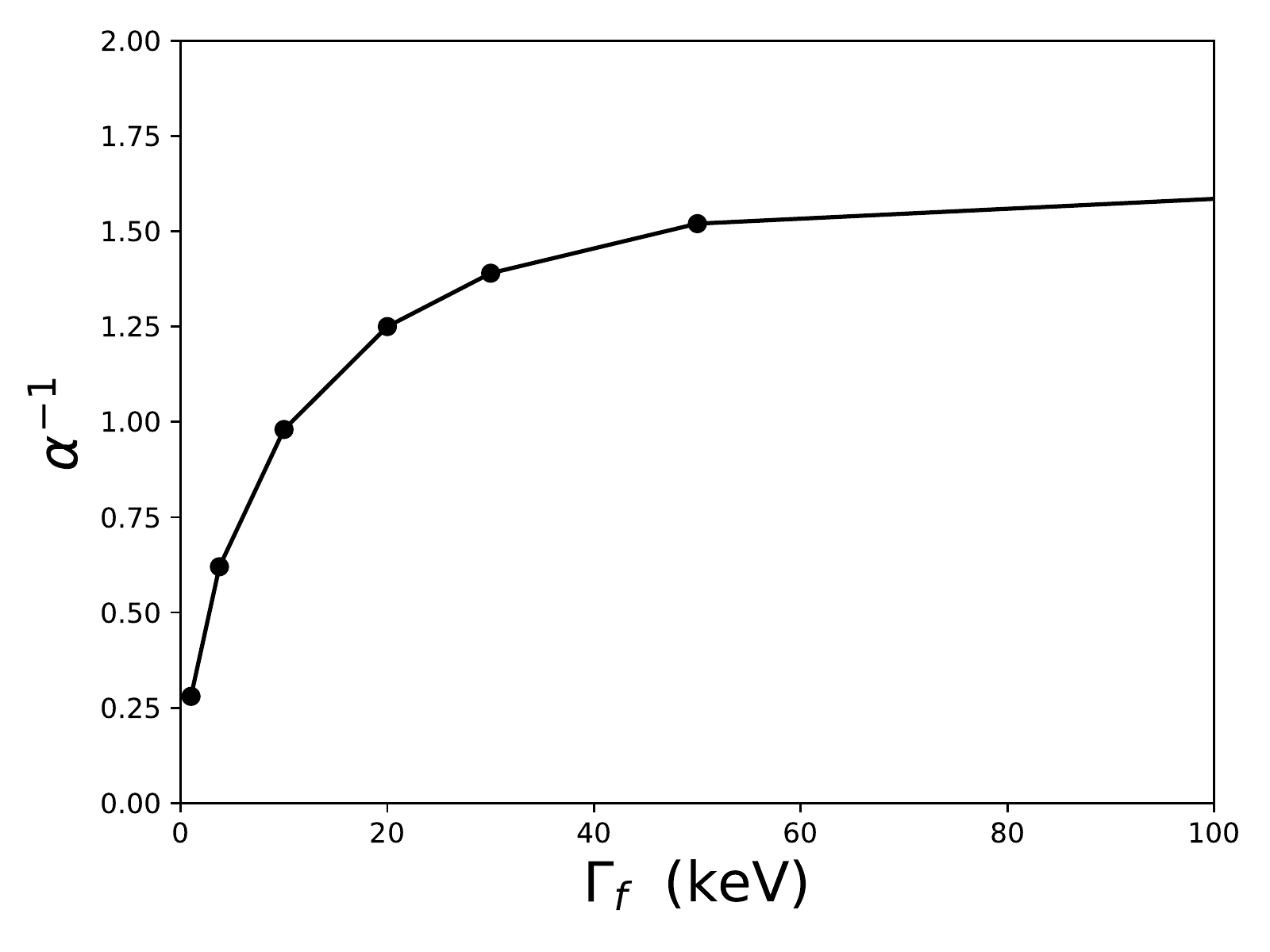}
\caption{
\label{insensitivity2}
Branching ratio $\alpha^{-1}$ for the full Hamiltonian 
as a function of fission width
$\Gamma_f$.
}
\end{figure}

We have also made a test of the 
insensitivity property  with the full Hamiltonian.  
Fig. \ref{insensitivity2} shows
the branching ratio as a function of the assumed fission 
width $\Gamma_f$ of the post-barrier reservoir states.  One
sees that the $\alpha^{-1}$ varies only by a factor of 1.5
over variation of $\Gamma_f$ by a factor of 20.  In short,
the branching ratio is largely determined by the probability to 
cross the bridge, rather than the decay rates on the far side.

\section{Diabatic interaction}

In this appendix we examine the diabatic interaction along
the $\Hb$ chain to confirm its systematic properties and
estimate its overall magnitude. They are calculated with
the code GCMaxial \cite{robledo} which evaluates 
the matrix elements by the Balian-Brezin formula \cite{ba69}.
The energy functional employed here is the Gogny D1S functional;
its PES was displayed in Fig. \ref{pes3}. For our application,
the PES configurations were obtained by the HF minimization procedure.
 We first demonstrate that Eq. (\ref{v_db}) offers
\begin{figure}
\includegraphics[width=1.0\columnwidth]{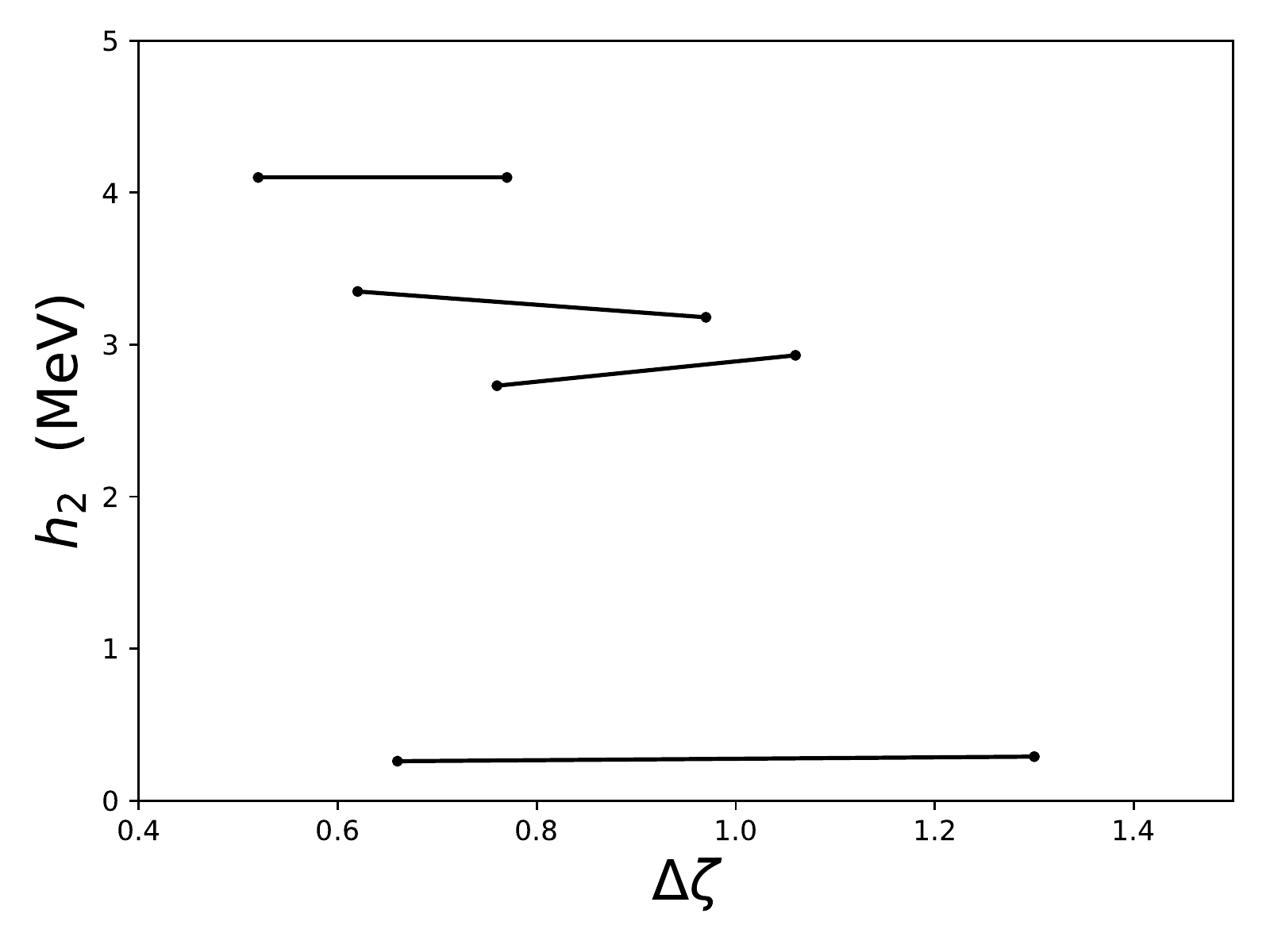}
\caption{
The diabatic matrix element $h_2$ 
given by Eq. (\ref{v_db}) 
for several sets of $(q_1,q_2)$ 
which satisfy $(q_1+ q_2)/2 = \bar Q  = 25$ b. 
The Gogny D1S funtional is employed. 
The quantity is plotted as a function of $\Delta \zeta$ for each set of 
$(q_1,q_2)$. 
\label{h2_vs_dz}
}
\end{figure}
a reasonable parameterization of the dependence on
$\zeta$, as was found in an early study \cite{bo90}.
In Fig. \ref{h2_vs_dz} the diabatic matrix element between configurations
at deformations  $(q_1+ q_2)/2 = \bar Q  = 25$ b are calculated as
a function of $q_1 - q_2$.  The plots show the derived
value of $h_2$ in Eq. (\ref{v_db}) as a function of $\Delta \zeta$. 
One sees that it is rather insensitive
to $\Delta \zeta$.  On the other hand, $h_2$ has a 
considerable variation among the different configurations.

\begin{figure}
\includegraphics[width=1.0\columnwidth]{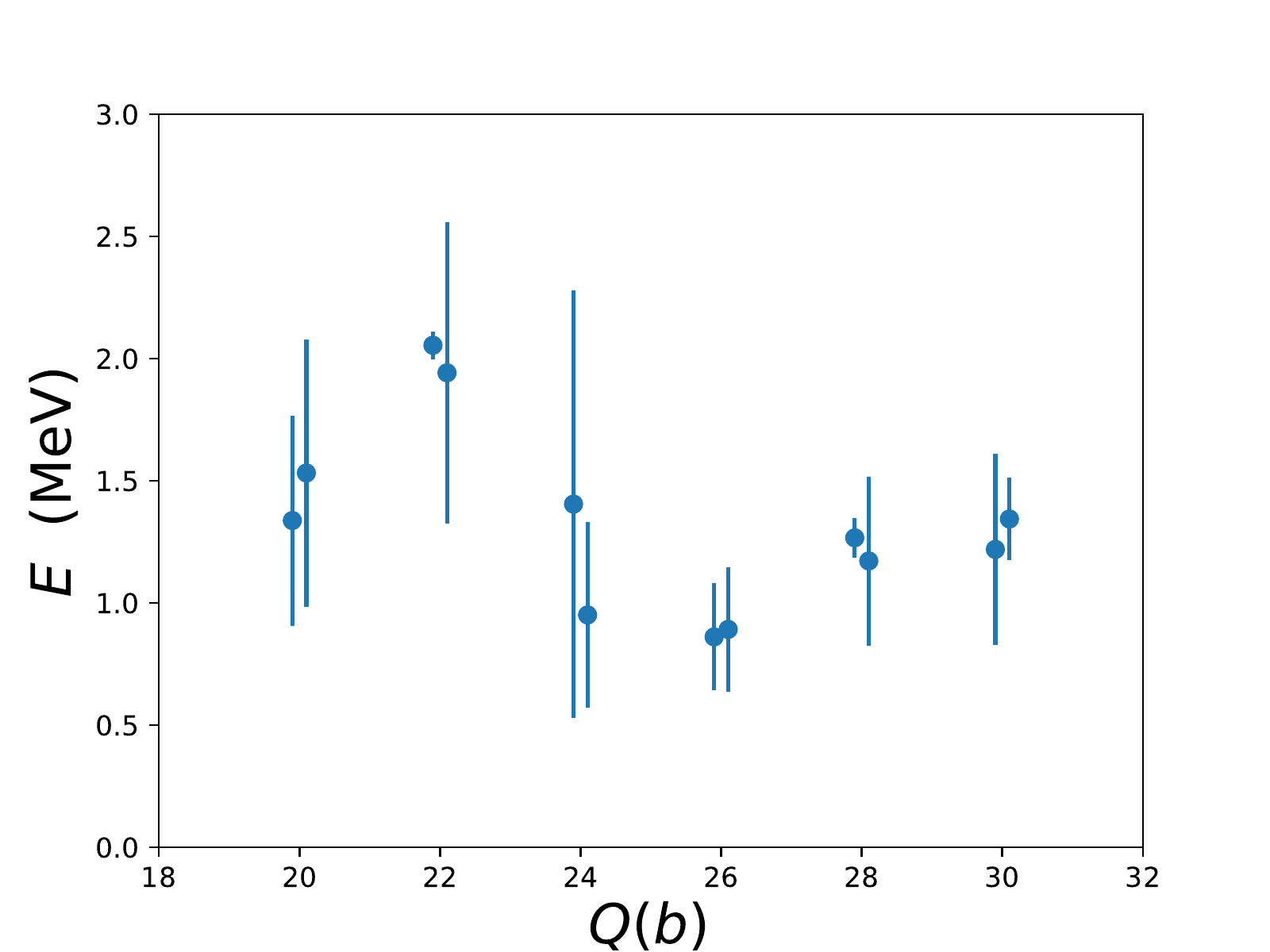}
\caption{
The average and the variance of $h_2$, evaluated by sampling 10 particle-hole 
configurations at each $\bar Q$. 
\label{diabatic-fig2}
}
\end{figure}

In this exploratory study we did not attempt to 
calculate these  matrix elements individually for each
diabatic link in the $\Hb$ chain.  Instead, we evaluated them for a sample
of configurations and used the average for constructing
$\Hb$.  Fig. \ref{diabatic-fig2}  shows the results for samples at
$\bar Q$ between 20 and 30 b, sampling 10 particle-hole
configurations at each point.  The dots show the averages
at each $\bar Q$ with the variance shown by the error bars.
The overall average $\langle i_{q_1}| H| i_{q_2} \rangle$ 
is about 1.5 MeV, and this is the value which we employed in the base parameters 
shown in Table II. 

\end{document}